\documentclass[12pt]{article} \usepackage{latexsym} \textwidth 15cm
\begin{document}

\title{{\bf HAWKING RADIATION AND \\ BLACK HOLE THERMODYNAMICS}
\thanks{Alberta-Thy-18-04, hep-th/0409024, review article solicited for
a celebratory Focus Issue on Relativity, ``Spacetime 100 Years
Later,'' to be published in New Journal of Physics.}}

\author{
Don N. Page
\thanks{Internet address:
don@phys.ualberta.ca}
\\
Institute for Theoretical Physics\\
Department of Physics, University of Alberta\\
Edmonton, Alberta, Canada T6G 2J1
}
\date{(2004 September 3; additional section added 2004 December 31
\thanks{On this date the author reached an age which is an integer
divisor of the number of references.}\ )}
\maketitle
\large

\begin{abstract}

\baselineskip 18 pt

An inexhaustive review of Hawking radiation and black hole
thermodynamics is given, focusing especially upon some of the
historical aspects as seen from the biased viewpoint of a minor player
in the field on and off for the past thirty years.

\end{abstract}
\normalsize

\baselineskip 18 pt
\newpage

\section{Historical Background}

Black holes are perhaps the most perfectly thermal objects in the
universe, and yet their thermal properties are not fully understood.
They are described very accurately by a small number of macroscopic
parameters (e.g., mass, angular momentum, and charge), but the
microscopic degrees of freedom that lead to their thermal behavior
have not yet been adequately identified.

Strong hints of the thermal properties of black holes came from the
behavior of their macroscopic properties that were formalized in the
(classical) four laws of black hole mechanics \cite{BCH}, which have
analogues in the corresponding four laws of thermodynamics:

The zeroth law of black hole mechanics is that the surface gravity
$\kappa$ of a stationary black hole is constant over its event horizon
\cite{Carter,BCH}.  This is analogous to the zeroth law of
thermodynamics, that the temperature $T$ is constant for a system in
thermal equilibrium.

The first law of black hole mechanics expresses the conservation of
energy by relating the change in the black hole mass $M$ to the
changes in its area $A$, angular momentum $J$, and electric charge $Q$
in the following way:
 \begin{equation}
 \delta M = {1\over 8\pi} \kappa \delta A + \Omega \delta J
    + \Phi \delta Q,
 \label{eq:1}
 \end{equation}
where an extended form of the zeroth law implies that not only the
surface gravity $\kappa$, but also the angular velocity $\Omega$ and
the electrostatic potential $\Phi$ are constant over the event horizon
of any stationary black hole.  This first law is essentially the same
as the first law of thermodynamics.

The second law of black hole mechanics is Hawking's area theorem
\cite{Hawkingarea}, that the area $A$ of a black hole horizon cannot
decrease.  This is obviously analogous to the second law of
thermodynamics, that the entropy $S$ of a closed system cannot
decrease.

The third law of black hole mechanics is that the surface gravity
$\kappa$ cannot be reduced to zero by any finite sequence of
operations \cite{Israel}.  This is analogous to the weaker (Nernst)
form of the third law of thermodynamics, that the temperature $T$ of a
system cannot be reduced to absolute zero in a finite number of
operations.  However, the classical third law of black hole mechanics
is not analogous to the stronger (Planck) form of the third law of
thermodynamics, that the entropy of a system goes to zero when the
temperature goes to zero.

Thus the four laws of black hole mechanics are analogous to the four
laws of thermodynamics if one makes an analogy between temperature $T$
and some multiple of the black hole surface gravity $\kappa$, and
between entropy $S$ and some inversely corresponding multiple of the
black hole area $A$.  That is, one might say that $T = \epsilon
\kappa$ and $S = \eta A$, with $8\pi\epsilon\eta = 1$, so that the
$\kappa \delta A/(8\pi)$ term in the first law of black hole mechanics
becomes the heat transfer term $T\delta S$ in the first law of
thermodynamics.

Even before the formulation of the four laws of black hole mechanics,
Bekenstein \cite{Bek1,Bek2,Bek3,Bek4} proposed that a black hole has
an entropy $S$ that is some finite multiple $\eta$ of its area $A$.
He was not able to determine the exact value of $\eta$, but he gave
heuristic arguments for conjecturing that it was $(\ln 2)/(8\pi)$ (in
Planck units, $\hbar = c = G = k = 4\pi\epsilon_0 = 1$, which I shall
use throughout).

However, for the first law of black hole mechanics to be equivalent to
the first law of thermodynamics, this would logically imply that the
black hole would have to have a temperature $T$ that is a
corresponding nonzero multiple of the surface gravity $\kappa$.  E.g.,
if $\eta = (\ln 2)/(8\pi)$ as Bekenstein proposed, then one would get
$\epsilon = 1/(\ln 2)$, so that $T = \kappa/(\ln 2)$.  But since it
was thought then that black holes can only absorb and never emit, it
seemed that black holes really would have zero temperature, or
$\epsilon = 0$, which would make Bekenstein's proposal inconsistent
with any finite $\eta$ \cite{BCH}.

Nevertheless, by a quite independent line of reasoning that was not
directly motivated by Bekenstein's proposal that he had rejected
\cite{BCH}, Hawking made the remarkable discovery that black holes are
not completely black but instead emit radiation \cite{Haw1,Haw2}.
Once he found that the radiation had a thermal spectrum, he realized
that it did make Bekenstein's idea consistent, of a finite black hole
entropy proportional to area, though not Bekenstein's conjectured
value for $\eta$.  In fact, Hawking found that the black hole
temperature was $T = \kappa/(2\pi)$, so $\epsilon = 1/(2\pi)$ and
hence $\eta = 1/4$.  This gives the famous Bekenstein-Hawking formula
for the entropy of a black hole:
 \begin{equation}
 S_{\mathrm{bh}} = S_{\mathrm{BH}} \equiv {1\over 4} A.
 \label{eq:2}
 \end{equation}
Here the subscript bh stands for ``black hole,'' and the subscript BH
stands for ``Bekenstein-Hawking.''

A precursor of Hawking's discovery of emission by black holes were the
calculations by Parker \cite{Par1,Par2,Par3,Par4} and Fulling
\cite{Full1,Full2} of particle creation by expanding universes, which
developed the concepts of Bogoliubov transformations \cite{Bog} in
time-dependent geometries that were later used by Hawking.  However, to
the best of my knowledge, it was a surprise to everyone, including
Hawking, that the emission from a black hole persisted even when the
black hole became effectively static.

Since I did my Ph.D. thesis \cite{Pagethesis} on ``Accretion into and
Emission from Black Holes'' (but missed the opportunity to use the more
catchy title, ``The Ins and Outs of Black Holes'') and had many
discussions about it with Hawking during the 1974-75 year that he spent
at Caltech, it may be of interest to quote part of an introductory
section from that thesis on what I then saw as the historical background
of black hole emission (and now add a few more comments within square
brackets):

\begin{quotation}
The first prediction of emission by a black hole was made by
Zel'dovich \cite{Zel1,Zel2}.  He pointed out on heuristic grounds that a
rotating black hole should amplify certain waves and there
should be an analogous quantum effect of spontaneous radiation of
energy and angular momentum.  Later Misner \cite{Mis} and Starobinsky
\cite{Star} confirmed the amplification by a Kerr hole of scalar waves
in the ``superradiant regime'' (where the angular velocity of the
wavefronts is lower than that of the waves), and Bekenstein
\cite{Beka} showed that amplification should occur for all kinds of
waves with positive energy density.  However, the quantum effect
predicted by Zel'dovich was not universally known, and in fact Larry
Ford at Princeton University and I independently rediscovered it.

The argument for this spontaneous radiation was that in a quantum
analysis the amplification of waves is stimulated emission of quanta,
so that even in the absence of incoming quanta one should get
spontaneous emission.  By using the relation between the Einstein
coefficients for spontaneous and stimulated emission, one can
calculate the spontaneous rate from the amplification factor, as
Starobinsky \cite{Star} noted, at least when the spontaneous emission
probability is much less than unity.

A problem arose for neutrinos in that Unruh \cite{Unruh73} showed that
their waves are never amplified.  This result violated Bekenstein's
conclusion and seemed to be a breakdown in the Hawking \cite{Hawkingarea}
area theorem.  The reason for the violation was traced to a negative
local energy density of the classical neutrino waves at the horizon.
However, Feynman suggested (unpublished) that the lack of
amplification might be due to the Pauli exclusion principle, so that
incident neutrinos suppress spontaneous emission which otherwise
occurs.  The amplification factor would then be less than unity, since
the calculation of an unquantized neutrino wave cannot directly show
the spontaneous emission but only how the emission changes as the
incident flux is varied.

[I had met up with Feynman at lunch at Caltech that day in early 1973
to ask him about my idea of spontaneous emission from rotating black
holes, just hours before I became somewhat crestfallen to find in
\cite{Zel1} that Zel'dovich had already pointed out this effect.
Feynman volunteered to come over, after stopping off to watch some
belly dancers on campus that day, to the offices of Kip Thorne's
graduate students Bill Press, Saul Teukolsky, and me to discuss, and
eventually agree with, my idea that at that time I still thought was
original.  When presented with the problem with the lack of
amplification for the classical neutrino waves, Feynman began drawing
diagrams on the blackboard, while noting, ``I'm supposed to be good at
these diagrams.'']

One might be surprised to find such a difference between integral and
half-integral spins showing up in the behavior of their unquantized
waves, but this is merely an illustration of the connection between
spin and statistics.  Pauli \cite{Pauli} has shown that half-integral
spins must be assigned anticommutation relations in order to get a
positive energy density, which is precisely what the unquantized
neutrino waves violate in not showing superradiance.

Indeed, this same behavior occurs in the Klein paradox.  A scalar wave
incident on an electrostatic potential step higher than the kinetic
energy plus twice the mass gives a reflected current greater than the
incident current.  On the other hand, a Dirac wave on such a step
gives less reflected current.  (This is the result if one makes the
causality requirement of the transmitted waves' having a group
velocity away from the step, rather than having the momentum vector
away from the step as in Bjorken and Drell \cite{BD}.)  Nikishov
\cite{Nik} uses field theory to calculate the pair production by a
potential step of general shape with no particles incident.  His
results show that the number of expected particles emitted in a given
Klein-paradox state is
 \begin{equation}
 \langle N \rangle = \pm (A-1),
 \label{III.6}
 \end{equation}
where $A$ is the amplification factor for the reflected wave of the
unquantized Klein-Gordon $(+)$ or Dirac $(-)$ equation.  This formula
applies even if the emission probabilities are not small, so that
$\langle N \rangle$ includes the possibility of emitting more than one
particle (if a boson) in the same state.

Unruh \cite{Unruh74} made a formal calculation of second quantization
of scalar and neutrino fields in the complete Kerr metric and found
essentially the same results as Eq. (\ref{III.6}) if he chose the
initial vacuum state to correspond to no particles coming out of the
past horizon.  Ford \cite{Ford} quantized the massive scalar field in
a somewhat different way with similar results.  However, Unruh noted
that the actual situation might be different, with no past horizon but
the black hole formed by collapse.  Nevertheless, neither he nor any
of the discoverers of the spontaneous emission attempted to calculate
that situation.

Meanwhile (summer 1973), Stephen Hawking at Cambridge University heard
of this work through Douglas Eardley [who as a postdoc at Caltech had
learned of it from discussions with Press, Teukolsky, and me] and so
while in Moscow discussed it with Zel'dovich and Starobinsky.
Believing in the reality of the spontaneous emission but wishing to
put its derivation on a firmer footing, Hawking dared to attempt the
difficult calculation of field theory during the collapse and
formation of a black hole.  Separating out the essential elements,
Hawking found how to calculate the particle emission at late times,
after the collapse had settled down to form a stationary black hole.
At first Hawking got an infinite number of particles emitted, but then
he discovered that the infinity corresponded to emission at a steady
rate.  However, the emission was not only in the superradiant states
or modes but in all modes that could come from the black hole!

Hawking initially did not believe this result (a consolation to those
of us who doubted it also when we first heard it).  Thinking that the
emission might be an artifact of the spherical symmetry he had
assumed, Hawking considered nonspherical collapse and got the same
emission.  Then he tried putting in a cutoff on the frequencies of the
modes in the initial state before the collapse, but that eliminated
all the emission, including the spontaneous emission in the
superradiant modes that Hawking was certain existed.  Perhaps most
convincing to Hawking was the fact that the emission rate was just
that of a thermal body with the same absorption probabilities as the
black hole and with a temperature (in geometrical units) equal to the
surface gravity of the hole divided by $2\pi$.  This result held for
fields of any spin and seemed to confirm some thermodynamic ideas of
Bekenstein \cite{Bek3}.  However, before the emission process was
discovered, Bardeen, Carter, and Hawking \cite{BCH} had argued against
Bekenstein's suggestion of a black-hole temperature proportional to
surface gravity.  Thus Bekenstein's ideas were originally not a
motivation for Hawking's calculation.

As word of his calculation began to spread, Hawking published a
simplified version of it in \emph{Nature} \cite{Haw1}.  However, even
at this stage Hawking was not certain of the result and so expressed
the title as a question, ``Black hole explosions?''  He noted that the
calculation ignored the change in the metric due to the particles
created and to quantum fluctuations.  One objection raised by several
people was that the calculation seemed to give a very high energy flux
just outside the horizon, which might prevent the black hole from
forming at all.  Hawking later answered this and other problems by a
more detailed version of the calculation \cite{Haw2}, which showed
that an infalling observer would not see many particles near the
horizon.  However, it might be noted that there is still some
controversy about the existence of particles there.  The back reaction
of the particles created would, in Hawking's view, simply be to reduce
the mass of the hole by the amount of the energy radiated away.

Presumably quantum fluctuations of the metric itself can give rise to
the emission of gravitons in addition to the emission of other
particles calculated as if the geometry were fixed.  By considering
linearized fluctuations in the metric about a given background, the
emission of gravitons can be handled in the same manner as the
emission of any other particles, though one might argue that graviton
emission depends more fundamentally upon the assumed fluctuations in
the metric.  Therefore, any observed consequences of graviton emission
can be viewed as testing whether gravity is quantized.

Hawking has argued (unpublished) that quantum mechanics allows small
deviations of the action from the extremum value that gives the
classical field equations for matter and geometry.  Thus the classical
equations can be violated in a small region near a black hole, giving
rise to the emission of matter or gravitational waves, but the
equations cannot be violated significantly on a very large surface
surrounding the hole.  Therefore, quantities determined by surface
fluxes at infinity do remain conserved: energy, momentum, angular
momentum, and charge.  This is the basis for arguing that the emission
carries away the quantities of the hole which otherwise would be
constant.  Note that baryon and lepton numbers are not observed to be
connected with long-range fields, so they presumably cannot be
determined by surface fluxes at infinity and thus would not be
conserved globally by the black-hole emission process.

The thermal emission first calculated by Hawking has been verified by
several subsequent calculations.  Boulware \cite{Boul} and Davies
\cite{Dav} have calculated the emission from a collapsing shell.
Gerlach \cite{Ger} has interpreted the emission as parametric
amplification of the zero-point oscillations of the field inside the
collapsing object.  DeWitt \cite{DeW} has given detailed derivations
of both the spontaneous emission process in the complete Kerr metric
(with no particles coming out of the past horizon) and of the thermal
emission from a black hole formed by collapse.  Unruh \cite{Unruh75}
has found that his derivation in the complete Kerr metric will give
not only the spontaneous but also the thermal emission if the boundary
condition at the past horizon is changed from no particles seen by an
observer at fixed radius just outside the horizon to no particles seen
by an observer freely falling along the horizon.  Wald \cite{Wald75},
Parker \cite{Park} and Hawking \cite{Haw3} have calculated the density
matrix of the emitted particles and find that it, as well as the
expected number in each mode, is precisely thermal.  [For a time
Hawking thought that particles escaped only in pairs, which led me to
come back from a spring camping trip in the snow in the Sierras in
1975 thinking I had figured out how to violate the second law of
thermodynamics.  Never have I made that error again.]  Bekenstein
\cite{Bek5a} has given an information-theory argument of why this
should be so.  Hartle and Hawking \cite{HH} have done a path-integral
calculation of the probability for a particle to propagate out of a
black hole from the future singularity and show that this method also
leads to the same thermal spectrum.  In summary, the thermal emission
from a black hole has been derived in a variety of ways by several
people, so its prediction seems to be a clear consequence of our
present theories of quantum mechanics and general relativity.

\end{quotation}

That of course was my own personal view as I was finishing my Ph.D. in
1976, heavily influenced by discussions with Hawking and with a few
others, but without in any way being a claim to a completely balanced
and broad view of what others might have been thinking at the time.
However, I thought it might be of at least some historical interest to
present here this biased viewpoint of the important historical
development of black hole emission.  For other viewpoints, see
\cite{Hawreview,Bekreview,Wheeler}.

\section{Hawking Emission Formulae}

For the Kerr-Newman metrics \cite{Kerr,Newman}, which are the unique
asymptotically flat stationary black holes in Einstein-Maxwell theory
\cite{Israeluniqueness,Carteruniqueness,HE,Robinson,Mazur}, one can
get explicit expressions for the area $A$, surface gravity $\kappa$,
angular velocity $\Omega$, and electrostatic potential $\Phi$ of the
black hole horizon in terms of the macroscopic conserved quantities of
the mass $M$, angular momentum $J \equiv Ma \equiv M^2 a_*$, and
charge $Q \equiv MQ_*$ of the hole \cite{Page76}, using the value
$r_+$ of the radial coordinate $r$ at the event horizon as an
auxiliary parameter:
\begin{eqnarray}
r_+ &=& M+(M^2-a^2-Q^2)^{1/2} = M[1+(1-a_*^2-Q_*^2)^{1/2}],
 \nonumber \\
A &=& 4\pi(r_+^2+a^2) = 4\pi M^2[2-Q_*^2+2(1-a_*^2-Q_*^2)^{1/2}],
 \nonumber \\
\kappa &=& {4\pi(r_+-M)\over A}
 = {1\over 2}M^{-1}[1+(1-{1\over 2}Q_*^2)(1-a_*^2-Q_*^2)^{-1/2}]^{-1},
 \nonumber \\
\Omega &=& {4\pi a\over A}
 = a_*M^{-1}[2-Q_*^2+2(1-a_*^2-Q_*^2)^{1/2}]^{-1},
 \nonumber \\
\Phi &=& {4\pi Q r_+ \over A}
 = Q_* {1+(1-a_*^2-Q_*^2)^{1/2} \over 2-Q_*^2+2(1-a_*^2-Q_*^2)^{1/2}}.
 \label{eq:3}
 \end{eqnarray}

Here $a_* = a/M = J/M^2$ and $Q_* = Q/M$ are the dimensionless angular
momentum and charge parameters in geometrical units
($G=c=k=4\pi\epsilon_0=1$ but for this without setting $\hbar=1$, so
that mass, time, length, and charge all have the same units, and
angular momentum has units of mass or length squared; e.g., the
angular momentum of the sun is $J_{\odot} \equiv a_{*\odot}M_{\odot}^2
= (0.2158 \pm 0.0017)M_{\odot}^2 = 47.05 \pm 0.37 \; \mathrm{hectares}
= 116 \pm 1 \; \mathrm{acres}$ \cite{Pij,MAIN}.  However, we shall
return to Planck units for the rest of this paper, so that every
quantity is dimensionless.)  For a nonrotating uncharged stationary
black hole (described by the Schwarzschild metric), $a_* = Q_* = 0$,
so $r_+ = 2M$, $A = 16\pi M^2$, $\kappa = M/r_+^2 = 1/(4M)$, $\Omega =
0$, and $\Phi = 0$.

Then Hawking's black hole emission calculation \cite{Haw1,Haw2} for
free fields gives the expected number of particles of the $j$th
species with charge $q_j$ emitted in a wave mode labeled by frequency or
energy $\omega$, spheroidal harmonic $l$, axial quantum number or
angular momentum $m$, and polarization or helicity $p$ as
 \begin{equation}
 N_{j\omega lmp} =
  \Gamma_{j\omega lmp}\{\exp[2\pi\kappa^{-1}(\omega-m\Omega-q_j\Phi)]
   \mp 1\}^{-1}.
 \label{eq:4}
 \end{equation}

Here the upper sign (minus above) is for bosons, and the lower sign
(plus above) is for fermions, and $\Gamma_{j\omega lmp}$ is the
absorption probability for an incoming wave of the mode being
considered.

More accurately, $\Gamma_{j\omega lmp}$ is the negative of the
fractional energy gain in a scattered classical wave with only inward
group velocity at the black hole horizon.  $\Gamma_{j\omega lmp}$ is
positive for all fermionic wave modes and for bosonic wave modes with
$\omega-m\Omega-q_j\Phi > 0$, which are at least partially absorbed by
the hole, but it is negative for bosonic superradiant modes with
$\omega-m\Omega-q_j\Phi < 0$, which are amplified by the hole.  Thus
$0 \le \Gamma_{j\omega lmp} \le 1$ for fermionic modes, but one just
has $\Gamma_{j\omega lmp} \le 1$ for bosonic modes, with
$\Gamma_{j\omega lmp}$ allowed to be negative for them.

Nevertheless, $N_{j\omega lmp}$ is never negative, because the thermal
Planck factor is also negative for bosonic superradiant modes.
$N_{j\omega lmp}$ also never diverges, even though the Planck factor
for bosons diverges as $\omega-m\Omega-q_j\Phi$ is taken to zero,
since then $\Gamma_{j\omega lmp}$ also goes to zero linearly with
$\omega-m\Omega-q_j\Phi$ and so keeps $N_{j\omega lmp}$ finite.
Then one can combine $\Gamma_{j\omega lmp} \le 1$ with Eq. (\ref{eq:4})
to get the double inequality
 \begin{equation}
 \mp N_{j\omega lmp} \le \Gamma_{j\omega lmp} \le 1.
 \label{eq:4b}
 \end{equation}

In the approximation of a stationary geometry with no back reaction,
the density matrix of the Hawking radiation is (for free fields) the
uncorrelated tensor product of thermal density matrices for each of
the modes with definite frequency, angular momentum, and charge.  The
thermal density matrices for each mode are diagonal in the number
basis, with the probability of $n$ particles in the mode being
 \begin{equation}
 P_n = N^n (1\pm N)^{-n\mp 1},
 \label{eq:5}
 \end{equation}
where for brevity I have here dropped the mode-labeling subscripts on
the expected number $N_{j\omega lmp}$ of particles in the mode.  Here
$n$ can be any nonnegative integer for bosons (upper sign) but is
restricted to be 0 or 1 for fermions (lower sign).

The von Neumann entropy for the thermal density matrix of each mode is
 \begin{equation}
 \delta S_{\mathrm{rad}} = - \sum_n P_n \ln P_n
              = (N \pm 1)\ln{(1 \pm N)} - N\ln{N}.
 \label{eq:5b}
 \end{equation}

Since the expected loss of energy, angular momentum, and charge of the
hole from emitting $N$ particles in the mode are $N\omega$, $Nm$, and
$Nq_j$ respectively, the expected change in the black hole entropy
from that emission mode is
 \begin{equation}
 \delta S_{\mathrm{bh}} = - N [2\pi\kappa^{-1}(\omega-m\Omega-q_j\Phi)]
  = - N \ln{\left({\Gamma \pm N \over N}\right)},
 \label{eq:5c}
 \end{equation}
where now I have omitted the mode-labeling subscripts not only on
$N_{j\omega lmp}$, but also on $\Gamma_{j\omega lmp}$.

Then the total expected change in the entropy of the world from the
emission of the mode in question is
 \begin{equation}
 \delta S = \delta S_{\mathrm{rad}} + \delta S_{\mathrm{bh}}
  = \pm \ln{(1\pm N)} 
      +N\ln{\left(1+{1-\Gamma \over \Gamma \pm N} \right)}
          \ge \pm \ln{(1\pm N)} \ge 0,
 \label{eq:5d}
 \end{equation}
with the extreme right inequality being saturated only if there is no
emission, $N=0$.

Thus the Hawking emission from a black hole into empty space obeys the
second law of thermodynamics, and it actually produces entropy from
all modes with nonzero emission.  This is as one would expect, since
the emission from a black hole with $T_{\mathrm{bh}} > 0$ into empty
space with $T=0$ is an out-of-equilibrium process.

It is important to note that since the expected number of particles $N
\equiv N_{j\omega lmp}$ depends not only on the Planck
factor but also on $\Gamma_{j\omega lmp}$, the effective temperature
$T_{j\omega lmp}$ varies from mode to mode.  The effective temperature
may be defined by the Boltzmann factor
 \begin{equation}
 {P_1 \over P_0} \equiv e^{-\omega/T_{j\omega lmp}}.
 \label{eq:8}
 \end{equation}
Since $P_1/P_0 = N/(1\pm N)$, one gets
 \begin{equation}
 T_{j\omega lmp} = 
  \omega/\ln{\left({1\pm N_{j\omega lmp}
    \over N_{j\omega lmp}}\right)}.
 \label{eq:8b}
 \end{equation}

When $m\Omega+q_j\Phi = 0$ (e.g., for the Schwarzschild metric, but also
for s-waves of neutral particles in any Kerr-Newman geometry), so that
the Planck factor becomes simply $1/(e^{2\pi\omega/\kappa} \mp 1)$,
and when $\Gamma_{j\omega lmp} = 1$, so that the classical incoming
wave is totally absorbed by the black hole, then $T_{j\omega lmp} =
T_{\mathrm{bh}} = \kappa/(2\pi)$, the Hawking temperature of the hole.
But otherwise, the effective temperature $T_{j\omega lmp}$ for the
mode generically depends on the mode.

For example, when $m\Omega+q_j\Phi = 0$, then generically
$\Gamma_{j\omega lmp} < 1$ and $T_{j\omega lmp} < T_{\mathrm{bh}}$.
For modes that have sufficiently large angular momentum in comparison
with their energy, so that they mostly miss the black hole and have
negligible absorption probability $\Gamma_{j\omega lmp}$, the
effective temperature is much less than the actual black hole
temperature.

On the other hand, for nearly extreme black holes, $1-a_*^2-Q_*^2 \ll
1$, which have low temperatures, $T_{\mathrm{bh}} \ll 1/(8\pi M)$, the
latter being the Schwarzschild value, and for modes with
$\omega-m\Omega-q_j\Phi < 0$ (both for bosonic superradiant modes with
$\Gamma_{j\omega lmp} < 0$ and for these fermionic modes that still
have $\Gamma_{j\omega lmp} > 0$), one can have $T_{j\omega lmp} \gg
T_{\mathrm{bh}}$.  This is the case in which the temperature of the
black hole has a negligible effect, and the Hawking emission formula
reduces approximately to Eq. (\ref{III.6}) above (where the
amplification factor is $A = - \Gamma_{j\omega lmp}$) for the
spontaneous emission first discovered by Zel'dovich \cite{Zel1,Zel2}.

From the mean number $N_{j\omega lmp}$ and the entropy $S_{j\omega
lmp}$ per mode, one can sum and integrate over modes to get the emission
rates of energy, angular momentum (the component parallel to the black
hole spin axis), charge, and entropy by the black hole:
 \begin{equation}
 {dE_{\mathrm{rad}}\over dt} = -{dM\over dt}
 = {1\over 2\pi}\sum_{j,l,m,p}\int \omega N_{j\omega lmp} d\omega,
 \label{eq:9}
 \end{equation}
 \begin{equation}
 {dJ_{\mathrm{rad}}\over dt} = -{dJ\over dt}
 = {1\over 2\pi}\sum_{j,l,m,p}\int m N_{j\omega lmp} d\omega,
 \label{eq:10}
 \end{equation}
 \begin{equation}
 {dQ_{\mathrm{rad}}\over dt} = -{dQ\over dt}
 = {1\over 2\pi}\sum_{j,l,m,p}\int q_j N_{j\omega lmp} d\omega,
 \label{eq:11}
 \end{equation}
 \begin{equation}
 {dS_{\mathrm{rad}}\over dt}
 = {1\over 2\pi}\sum_{j,l,m,p}\int S_{j\omega lmp} d\omega.
 \label{eq:12}
 \end{equation}
Here $M$, $J$, and $Q$ (without subscripts) denote the black hole's
energy, angular momentum, and charge.  By the conservation of the
total energy, angular momentum, and charge, the black hole loses these
quantities at the same rates that the radiation gains them.

This is not so for the total entropy, which generically increases, as
noted above.  The black hole entropy changes at the rate
 \begin{equation}
 {dS_{\mathrm{bh}}\over dt}
 = {1\over 2\pi}\sum_{j,l,m,p}\int
 [2\pi\kappa^{-1}(\omega-m\Omega-q_j\Phi)] N_{j\omega lmp} d\omega,
 \label{eq:13}
 \end{equation}
and by using Eq. (\ref{eq:4}), as in the derivation of
Eq. (\ref{eq:5d}), one can show that the total entropy
$S=S_{\mathrm{bh}}+S_{\mathrm{rad}}$ (black hole plus radiation) changes
at the rate
 \begin{equation}
 {dS\over dt}
 = {1\over 2\pi}\sum_{j,l,m,p}\int d\omega
 \left[ \pm \ln{(1\pm N_{j\omega lmp})}
   +N_{j\omega lmp}\ln{\left(1+{1-\Gamma_{j\omega lmp} \over \Gamma_{j\omega lmp}
     \pm N_{j\omega lmp}} \right)} \right].
 \label{eq:14}
 \end{equation}

For the emission of $n_s$ species of two-polarization massless
particles of spin $s$ from a Schwarzschild black hole
(nonrotating and uncharged) into empty space, numerical calculations
\cite{Page76,Page76b,Page83} gave
 \begin{equation}
 {dE_{\mathrm{rad}}\over dt} = -{dM\over dt}
 = 10^{-5} M^{-2}(8.1830 n_{1/2} + 3.3638 n_1 + 0.3836 n_2),
 \label{eq:15}
 \end{equation}
 \begin{equation}
 {dS_{\mathrm{rad}}\over dt}
 = 10^{-3} M^{-1}(3.3710 n_{1/2} + 1.2684 n_1 + 0.1300 n_2),
 \label{eq:16}
 \end{equation}
 \begin{equation}
 {dS_{\mathrm{bh}}\over dt}
 = - 10^{-3} M^{-1}(2.0566 n_{1/2} + 0.8454 n_1 + 0.0964 n_2).
 \label{eq:17}
 \end{equation}

Therefore, if a Schwarzschild black hole emitted just massless
neutrinos into empty space, the entropy in the radiation would be
1.6391 times as much as the entropy decrease of the black hole; if it
emitted just photons, the radiation entropy would be 1.5003 times that
by which the hole decreased; if it emitted just gravitons, the
external entropy would be 1.3481 times the entropy drawn out of the
hole; and if three massless neutrino species were emitted along with
photons and gravitons (and negligible other particles), the radiation
entropy would be larger by a factor of 1.6187 \cite{Page83}.

\section{The Generalized Second Law}

Even if a black hole is not emitting into empty space, there are
strong arguments that the total entropy of the black hole plus its
environment cannot decrease.  This is the {\it Generalized Second Law}
(GSL).  Bekenstein first conjectured it when he proposed that black
holes have finite entropy proportional to their area
\cite{Bek1,Bek2,Bek3,Bek4}, and he gave various arguments on its
behalf, though it would have been violated by immersing a black hole
in a heat bath of sufficiently low temperature if the black hole could
not emit radiation \cite{BCH}.

Once Hawking found that black holes radiate \cite{Haw1,Haw2}, he
showed that the GSL held for a black hole immersed in a heat bath of
arbitrary temperature, assuming that the radiation thermalized to the
temperature of the heat bath.  Zurek and Thorne \cite{ZT}, and Thorne,
Zurek, and Price \cite{TZP}, gave more general arguments for the GSL
without this last assumption.  Their arguments were later fleshed out
in a mathematical proof of the GSL for any process involving a
quasistationary semiclassical black hole \cite{FP}.  Other proofs of
the GSL have also been given \cite{Sor1,Sor2,Waldbook,Sor3}.

With some exceptions \cite{Sor2,Sor3}, these proofs so far generally
have two key assumptions: (1) The black hole is assumed to be
quasistationary, changing only slowly during its interaction with an
environment.  It has been conjectured \cite{TZP} that the GSL also
holds, using the Bekenstein-Hawking $A/4$ formula for the black hole
entropy, even for rapid changes in the black hole, but this has not
been rigorously proved.  Even to make this conjecture precise would
require a precise definition of the entropy of the environment, which
is problematic in quantum field theory when one attempts to define the
entropy of quantum fields in some partial region of space (e.g., the
region outside the black hole) with a sharp boundary \cite{BKLS,Sred}.

(2) The semiclassical approximation holds, so that the black hole is
described by a classical metric which responds only to some average or
expectation value of the quantum stress-energy tensor.  This allows
the black hole entropy to be represented by $A/4$ of its classical
horizon.  This approximation also implies that the radiation from the
hole is essentially thermal, with negligible correlations between what
is emitted early and late in the radiation, so that one may use the
von Neumann entropy $S_{\mathrm{rad}} = -{\mathrm{tr}}(\rho\ln\rho)$
for the entropy of the radiation and yet have it plus $A/4$ for the
black hole to continue to increase (once a suitable way is chosen to
regularize the divergence of $-{\mathrm{tr}}(\rho\ln\rho)$ that one
would get from a sharp black hole boundary \cite{BKLS,Sred}).

Now if information is really lost down a black hole as Hawking
originally proposed \cite{Haw76b}, and if the Hawking radiation really
has negligible correlations between what is emitted early and late,
then it might be true that $A/4 -{\mathrm{tr}}(\rho\ln\rho)$, suitably
regularized, would never decrease.  But since this information loss
proposal has been controversial since near the beginning
\cite{Page80,Pagereview}, and since now even Hawking has given it up
\cite{HawGR17}, it might well be that information is not lost forever
down a black hole but instead comes back out with the radiation.  If
so, for a black hole formed from matter in nearly a pure state, the
total radiation from the hole (after it evaporates completely, as I
assume it will) will also be in nearly a pure state.  In this case,
when the black hole originally formed, the total entropy in the GSL
would be somewhat greater than $A/4$ and hence large.  However, after
the evaporation, there would be no black hole entropy, and the
radiation, in nearly a pure state, would have very little entropy.
Thus the total GSL entropy would have gone down enormously.

Of course, this same problem could arise in the second law for any
other composite system if the entropy were taken to be the sum of the
von Neumann entropies for each of the subsystems.  For example, if one
had a lump of coal in nearly a pure state with incoming radiation also
in nearly a pure state, the radiation could heat up the coal, which
would then radiate nearly thermally.  When the hot coal had only
partially cooled, there would be sufficient correlations between what
the coal had radiated and its own internal state that the von Neumann
entropy of both would be large.  Thus, by this coarse-grained
procedure of calculating the entropy of the coal plus the radiation
(by ignoring the entanglement or quantum correlations between these
two subsystems and just adding the von Neumann entropies of the
density matrices of each that are obtained by tracing over the rest of
the total system), one would get a high entropy at intermediate times.
However, at late times, the coal would cool back down to nearly its
ground state of low entropy, and the radiation would similarly be
nearly a pure state (with a large amount of subtle correlations
between its different parts), so the total von Neumann entropy of these
two parts would have decreased back down to a small value again.

This phenomenon illustrates the problem that nontrivial versions of
the second law usually require coarse graining, but then the result
depends on the coarse graining and may not always have the desired
property.  If one uses the von Neumann entropy of an entire closed
system with no coarse graining, then if this system evolved unitarily
(e.g., with no loss of information), then the von Neumann entropy is
simply a constant, and the second law becomes trivial.  Dividing a
total system up into subsystems, calculating the density matrix and
von Neumann entropy of each part by tracing over the rest of the
system, and then adding up the resulting entropies of each part
usually does give a nontrivial entropy by this coarse graining that
ignores the quantum correlations or entanglements between the
subsystems.  This nontrivial entropy can indeed increase if
correlations between the subsystems grow, so that more of the quantum
information about the total system goes into the correlations that are
ignored in this particular coarse-grained method of calculating the
entropy.

Typically in our universe spatially separated subsystems have less
than maximal correlations between them, and typically interactions
between these subsystems cause the correlations to grow with time.  In
this way the coarse graining that ignores these correlations gives an
increasing entropy and expresses the second law of thermodynamics for
our universe.  However, there can be exceptions, such as the coal that
cools off so that it no longer has the energy to remain significantly
correlated with the radiation it emitted.  (In this example one could
save the second law by dividing the radiation itself up into
subsystems whose correlations do not decrease with time, but this
example illustrates that the validity of this formulation of the
second law depends on the choice of coarse graining and may be
violated for certain choices.)

Since there are these problems with ordinary systems in giving a
precise nontrivial definition of entropy that always obeys the second
law of thermodynamics, one should not be too surprised that there may
also be problems with formulations of the second law for systems
containing black holes.  Therefore, it is probably unrealistic to
expect that there can be a rigorous proof of the second law (or of a
GSL) for black hole systems in all generality.

Nevertheless, we would expect that if we have a sufficient amount of
coarse graining (such as coarse graining the radiation or other black
hole environment into sufficiently many parts and ignoring their
correlations, as well as ignoring the correlations between the
radiation and the black hole), the GSL should almost always be valid.
This people have tested with a wide variety of gedanken experiments.

Before Hawking had discovered that black holes radiate, Bekenstein
\cite{Bek3,Bek4} realized that his GSL might be violated if an
entropy-carrying object could be lowered sufficiently near a black
hole (so that nearly all of its energy could be extracted first) and
then dropped in with its energy so low that the increase of the black
hole entropy would not balance the loss of the entropy of the object.
To avoid this violation of the GSL, Bekenstein proposed that there was
a limit on how close to the black hole an object with fixed entropy
and fixed local energy could be lowered.  This led Bekenstein to
conjecture \cite{Bek5} that the entropy $S$ of a system of energy $E$
and linear size $R$ was limited by the formula
 \begin{equation}
 S \le S_B(E,R) \equiv 2\pi ER.
 \label{eq:19}
 \end{equation}

This was a very interesting proposal in its own right, but it
developed that there are a lot of problems with it
\cite{Unwin,Page82,UW,Deutsch,UW2,Zas,PW,Page00a,Page00b,Page00c,MS}.
Perhaps the main difficulty is how to give precise definitions for the
system and for its $S$, $E$, and $R$ \cite{Page00c}.  For various
choices of those definitions, one could easily come up with
counter-examples to the conjecture.

For example, if the system is arbitrary quantum fields in some bounded
region of space, and if zero-point energies are not counted in $E$ (or
else one could violate the bound by negative Casimir energies
\cite{Unwin}), then by making the number of fields $N$ sufficiently
large, $S$ grows as $\ln{N}$ for fixed $E$ and $R$, allowing the bound
to be violated \cite{Page82,UW}.  Or, if $E$ is the expectation value
of the energy (over the ground-state value), then by taking a density
matrix formed from the ground state and a tiny mixture of an excited
state with small probability $p$, $E$ goes linearly with $p$, but $S$
goes as $p\ln{(1/p)}$, so $S/E \sim \ln{(1/p)}$, which diverges as $p
\rightarrow 0$ and is not bounded by $2\pi R$
\cite{Deutsch,Page00a,Page00b,Page00c}.  Also, if the bounded region
of space is sufficiently nonspherical, one can get violations with
large $S \gg 1 \gg S_B \equiv 2\pi ER$ even with small $N$
\cite{Page00a,Page00b}.

Bekenstein has given rebuttals \cite{Bek6,Bek7,Bek8,Bek9,Bek10} that
the system should be a complete system with positive energy, so that
presumably $S_B$ is not to be allowed to become arbitrarily small.  In
that case, in which $S_B$ is larger than some number that depends on
the number of fields $N$ etc., it does seem plausible that the actual
entropy $S$ may be bounded by $S_B$.  For example, for thermal
radiation in 4-dimensional spacetime, for $S_B \gg 1$ one gets $S \sim
S_B^{3/4}$ (with a coefficient depending on the number of fields
etc.), which is certainly bounded by $S_B$ for sufficiently large
$S_B$.  However, I would be skeptical that there is any reasonable
definition of $S$, $E$, and $R$ that allows $S_B \equiv 2\pi ER$ to be
arbitrarily small and yet maintains the bound $S \le S_B$.

Furthermore, if one follows Bekenstein's philosophy of considering
only complete systems of presumably bounded energy and momentum, it is
hard to see how to give a precise definition of the size $R$ that
would be finite if it encompassed the entire system, because of the
position-momentum uncertainty principle.  But if one takes a
definition of $R$ that makes it infinite, then $S_B$ becomes infinite,
and Bekenstein's conjectured bound becomes trivial.

I have proposed a definition of systems with finite $R$ that are
``vacuum outside $R$'' \cite{Page00c}, but then when $S_B$ is made
arbitrarily small and one uses the von Neumann definition for $S$, $S
> S_B,$ so Bekenstein's conjectured bound is violated.  One might want
to use instead a microcanonical definition of entropy, but that is
difficult to do for a finite-size complete system (even if just
``vacuum outside $R$''), since then the system cannot be composed of
any finite number of energy eigenstates and hence could not be
ascribed any finite microcanonical entropy.

For further work on Bekenstein's conjectured bound and its relation to
Raphael Bousso's covariant entropy bound
\cite{Bousso99,Bousso00,FMW,Bousso02,BFM}, see
\cite{Bousso01,Bousso03,Bousso04a,Bousso04b,Bousso04c}.

However, here the question is to what extent Bekenstein's proposed
bound is related to the Generalized Second Law.  Unruh and Wald and
others \cite{UW,UW2,Zas,PW} have argued that what saves the GSL is not
Bekenstein's proposed bound, but the buoyancy or flotations effect of
Hawking radiation, which prevents one from lowering an object close
enough to the black hole that one can extract enough energy from it to
give a violation of the GSL when the object falls in.

(As an historical aside, after the Unruh-Wald buoyancy mechanism was
published in 1982, I recalled privately proposing to Wald in early 1976
this mechanism for saving the GSL, but in 1976 it was not believed that
the Hawking radiation near the black hole horizon would really be
observable or have any significant buoyancy effect, so my proposal
seemed untenable and was dropped from consideration at that time. Wald
more recently told me he did not remember my suggestion when he and
Unruh independently rediscovered this mechanism after realizing that
the Unruh acceleration radiation \cite{Unracc} would make the buoyancy
effect real.  I have no reason at all to doubt his honesty about this,
especially since I myself did not remember and publish my own abandoned
suggestion even when it became apparent that there would be a real
buoyancy effect from the Unruh acceleration radiation. Perhaps the
moral of this incident is that even if there is an apparently strong
objection to your otherwise good idea, don't dismiss it too completely
from your memory.)

Bekenstein \cite{Bek6,Bek7,Bek8} has disputed the claim that the
buoyancy effect saves the GSL even without his conjectured entropy
bound.  However, it does seem to be the case that there are several
``proofs'' of the GSL that do not obviously require assuming
Bekenstein's proposed bound, so it seems that surely it is unnecessary
(though this argumentation does not rule out the possibility that some
form of Bekenstein's proposed bound might follow from some of the same
assumptions that implicitly go into the GSL, so that it really is
necessary, perhaps somewhat analogous to the way that $2+2=4$ is
logically necessary for Einstein's equations to follow from the
Einstein-Hilbert action, even though one may not need explicitly to
invoke $2+2=4$ in deriving Einstein's equations).

\section{Microscopic Description of Black Hole Entropy}

Even if it turns out that the Generalized Second Law is generally
valid under suitable circumstances, there is still the question of
what the entropy of a black hole represents.  For ordinary
thermodynamic systems, the entropy is in some sense roughly (the
logarithm of) a count of the number of states accessible to the
system.  That is, if a system has equal probabilities to be in any of
$N$ states (and no probability to be in any other states), then its
entropy is $\ln{N}$.  Of course, in general, the nonzero probabilities
are not all equal, but if the $n$th state has probability $p_n$, one
can say that in some crude sense it corresponds to $N_n = 1/p_n$
states, and then the entropy is the expectation value of $\ln{N_n}$,
using the probabilities $p_n$ as weights in averaging $\ln{N_n}$ over
all states $n$.

So for a black hole, the question is what all the
$\exp{S_{\mathrm{bh}}} \sim \exp{(A/4)}$ accessible states are.  Or,
to put it another way, what and where are the degrees of freedom of a
black hole?

One idea is that the degrees of freedom exist inside the black hole,
say in the matter that has fallen in and/or in the antiparticles
produced along with the particles emitted by the Hawking radiation
\cite{FN,BFZ}.  This is perhaps the simplest view, but it does leave
it difficult then to explain how the information about those degrees
of freedom can get out when the black hole evaporates, if indeed
information is not lost in black hole formation and evaporation.  A
possible way to resolve this difficulty is to say that quantum gravity
effects dissolve the absolute distinction between the inside and the
outside of the black hole, so that information that in a semiclassical
approximation appears to be forever hidden inside a black hole can
actually come out.  Perhaps there are quantum amplitudes for wormholes
from the ``inside'' to the outside, or perhaps just for tubular
regions or conduits of trivial topology where the causal structure is
sufficiently altered for the information to be conducted out
\cite{Pagereview}.  It seems likely that there must be some amplitudes
for such structures to occur, though the challenge would be to explain
how they can funnel {\it all} of the information back out by the time
that the black hole completely evaporates.

Another idea is that the degrees of freedom exist precisely on the
surface of the black hole, say in its shape \cite{Sor1,Sor3}.  There
have been calculations from various approaches to quantum gravity that
have counted the degrees of freedom of the horizon and have given
(perhaps not surprisingly) an entropy proportional to the surface area
\cite{Car95,Rovelli,Teitelboim,Car97,Rovelli2,Rovelli3,ABCK,AK,Car99,Car02}.
One counter-intuitive aspect of this idea is that locally there is
nothing special about the horizon (except perhaps when the geometry is
eternally stationary, so that the event horizon coincides with an
apparent horizon that can be located by quasi-local measurements of
the geometry).

A third idea is that the important degrees of freedom are just outside
the black hole horizon \cite{tHooft,BKLS,MI}.  This is supported by
calculations of the thermal atmosphere of a black hole, which is
believed to have real observable effects for the buoyancy of
hypothetical highly reflecting objects lowered extremely close to the
black hole \cite{UW}.  On the other hand, the thermal atmosphere has a
negligible effect on observers freely falling through the event
horizon, so this makes it somewhat difficult to believe that it really
would have the huge entropy that the black hole has.  Another problem
with ascribing the entropy to the thermal atmosphere is that a
semiclassical calculation of the entropy of quantum fields outside the
horizon of a classical black hole gives a divergent result, unless one
puts a cutoff on the modes close to the horizon, and then the
resulting entropy depends sensitively on the cutoff.

However, it seems there must be something to the argument that entropy
resides in the thermal atmosphere, since if one puts in a reflecting
boundary to exclude this thermal atmosphere above some height above
the horizon, then the total black hole entropy is reduced below $A/4$
by the amount one would ascribe to the thermal atmosphere that is
excluded by the boundary \cite{PageA4}.  If one could get the boundary
down to within about one Planck length of the horizon, then the
semiclassical calculation would say that the total entropy would be
reduced to zero.  Thus it is conceivable that all of the black hole
entropy resides in the thermal atmosphere, but since the semiclassical
approximation would break down if the boundary were placed that close
to the horizon, we cannot yet be sure.

Yet another idea, or set of ideas, is that one simply cannot localize
the degrees of freedom that give the black hole entropy.  This would
certainly seem to be the case in string/M theory, since the strings
and branes that are fundamental to that theory are nonlocal objects.

For example, one of the great successes of string/M theory is giving a
precise account (including Hawking's factor of 1/4 in $S_{\mathrm{BH}}
= A/4$) of the entropy of certain kinds of black holes in terms of an
extrapolation from certain D-brane configurations
\cite{Sen,SV,CM,HS,BMPV,JKM,MSt,BLMPSV,HMS,DVV,HLM,MSuss,KT,HRS,CT,
Mald,MStrom,HM,PolTASI,HP,MalStr,Vafa,Mald2,BFKS,MSW,SS,MAdSCFT,
Strom,Peet,BSS,Witten,GKS,Sen98,Cvetic98,Witten98,Skenderis99,
Gubser98,AGMOO,Marolf98,Horowitz99,Damour99,Khuri0,Peet0,DM,
Myers1,Khuri1,Mathur4,Thorlacius4,Duff4}.  These D-branes are
nonlocal.  However, it must be admitted that these calculations do not
fully elucidate the nature of the nonlocality in the black hole case,
because of the extrapolation needed from an understandable weak-field
D-brane configuration to a strong-field black hole configuration.

A related string/M theory picture of the degrees of freedom of a black
hole is that they reside in open strings attached to the horizon
\cite{Susskind}.  Matter falling through the horizon may be
represented by closed strings approaching the black hole that become
open strings attached to the horizon.  Then these open strings
interact strongly and eventually go back into closed strings being
emitted from the horizon as Hawking radiation.  Because the open
strings on the horizon retain the information brought in during the
black hole formation (until they are radiated back out), information
is not lost in this picture.  However, the strong interactions between
the strings attached to the horizon mean that the information is
highly scrambled or encoded, so that what is radiated can look very
nearly like thermal radiation.

This is at least the view for observers that stay outside the black
hole.  For observers that fall through the horizon, they do not see
anything special about that surface, and they see the information
falling through the horizon.  It would then seem that the information
must have been cloned, so that a copy of what falls into the black
hole is retained on its surface.  However, cloning of information is
forbidden in linear quantum theory \cite{clone}, so this raises a
puzzle.  Building upon some ideas of 'tHooft
\cite{tHooft,'tHooft2,'tHooft3}, Susskind and collaborators
\cite{STU,Suss,ST,Suss2,LSU,LPSTU,SU} have proposed the principle of
{\it Black Hole Complementarity}, that it doesn't matter that copies
of the information have been made, since no single observer can access
more than one copy.

Perhaps a related way to give an heuristic justification of Black Hole
Complementarity is the following argument:  In ordinary quantum field
theory in a fixed classical globally hyperbolic spacetime, information
cannot be cloned to appear twice on some (spatial) Cauchy
hypersurface.  However, unitary evolution of the quantum fields means
that the same information actually does occur at different times (the
same information on all Cauchy surfaces).  Therefore, if one has a
surface that is not everywhere spacelike, it can be connected to
itself by causal curves through the spacetime and can have the same
information appearing twice, say within any two regions on the surface
that are connected by causal curves through the spacetime.  Now in the
black hole spacetime, the hypersurface where the information is
supposed to appear twice (once outside the black hole in the form of
Hawking radiation to be seen by an outside observer, and once inside
to be seen by an infalling observer) has its normals highly boosted
from one region to another, so in a sense the hypersurface becomes
nearly null.  Then if there are quantum uncertainties in the
four-geometry, it may be indefinite whether or not the surface is
really a spacelike surface and therefore whether there really is a
problem with having the same information appearing twice on it.

That is, in quantum gravity, one would not expect a definite
four-metric or even a definite causal structure, so that one cannot
say with definiteness which regions are not causally related and which
therefore cannot be given copies of the same information.  In
particular, one may never be able to say with precision that two
operators in two different regions commute (or anti-commute), because
one cannot say with precision that the two regions are spacelike
separated (are not causally connected through the quantum spacetime).
(As an aside, it would seem to me that this might lead to difficulties
in canonical quantum gravity and the Wheeler-DeWitt equation, in which
one attempts to write a quantum state as a functional of the
three-geometry and matter configuration on some three-surface, which
seems to assume implicitly that the local geometry and matter field
variables commute for different regions of the three-surface.)

Because it seems that the degrees of freedom describing a black hole
cannot be localized, and since they presumably cannot be described
even in terms of some four-geometry with a definite causal structure,
it may be difficult to try to give much of a description of them until
we have and understand a good theory of quantum gravity.

\section{Logarithmic corrections to black hole entropy}

One area of recent developments that has not yet settled down to
definitive conclusions concerns corrections to the Bekenstein-Hawking
formula (\ref{eq:2}) that equates the black hole entropy
$S_{\mathrm{bh}}$ to the Bekenstein-Hawking expression $S_{\mathrm{BH}}
\equiv A/4$, one-fourth the area of the event horizon.  Since this
formula was derived by Hawking from his semiclassical calculation of
the black hole temperature, it would be expected to have quantum
corrections.

One type of correction would come from the one-loop effects of quantum
matter fields near a black hole.  For example, Fursaev \cite{Fur}
calculated that with $N_s$ massless fields of spin $s$ present, the
entropy of a Schwarzschild black hole with $S_{\mathrm{BH}} \equiv A/4
\gg 1$ would be
 \begin{equation}
 S_{\mathrm{bh}} = S_{\mathrm{BH}} + {1\over 360}(4N_0 + 7N_{1/2} -
 52N_1 - 233N_{3/2} + 848N_2) \ln{S_{\mathrm{BH}}} + O(1).
 \label{eq:20}
 \end{equation}

Similarly, Mann and Solodukhin \cite{ManSol} calculated that one-loop
effects of a single scalar field would modify the entropy of a large
extreme Reissner-Nordstrom black hole to give
 \begin{equation}
 S_{\mathrm{bh}} = S_{\mathrm{BH}} - {1\over 180} \ln{S_{\mathrm{BH}}} +
 O(1).
 \label{eq:21}
 \end{equation}
Note that for the Reissner-Nordstrom black hole, the coefficient of the
logarithmic term is $-1/2$ what Fursaev calculated it to be for the
Schwarzschild black hole with a single scalar field.

However, the main emphasis recently has been on purely quantum gravity
corrections to the entropy.  Kaul and Majumdar \cite{KM} performed a
quantum geometry calculation that gave, again for large
$S_{\mathrm{BH}}$,
 \begin{equation}
 S_{\mathrm{bh}} = S_{\mathrm{BH}} - {3\over 2} \ln{S_{\mathrm{BH}}} +
 O(1).
 \label{eq:22}
 \end{equation}
Note that the nonperturbative calculation of Kaul and Majumdar gives the
coefficient of the logarithm as $-135/212$ times what one would get from
the result of the one-loop calculation of Fursaev with one single spin-2
particle (e.g., the graviton).

Carlip \cite{Carlip} reproduced the Kaul and Majumdar result from
logarithmic corrections to the Cardy formula \cite{Cardy,BCN}, and
Govindarajan, Kaul, and Suneeta \cite{GKS2} also found it from an exact
expression for the partition function of the Euclidean BTZ black hole
\cite{BTZ,BHTZ}.  Rama \cite{Rama} did a calculation of the asymptotic
density of open $p$-brane states and found a logarithmic term with a
coefficient $-(p+1)/(2p)$, which agrees with the Kaul-Majumdar
coefficient above for open strings ($p=1$).  Gour \cite{Gour} obtained
the Kaul-Majumdar result for a particular approach to black hole
quantization.

However, more recently there have been other calculations
\cite{DL,Meiss,GM1,GM2} that have given the coefficient of the
logarithm as $-1/2$ rather than $-3/2$, so that
 \begin{equation}
 S_{\mathrm{bh}} = S_{\mathrm{BH}} - {1\over 2} \ln{S_{\mathrm{BH}}} +
 O(1).
 \label{eq:23}
 \end{equation}
 
Other papers on logarithmic corrections for black hole entropy include
\cite{GM94,DKM,DMB,MP,NOO,CG,CM2,Set1,Set2,
ACAP,Hod,Myung,Park2,AL,Med1,CM3,More,Med2,Khrip}.

It should be noted that there are different ways to define the black
hole entropy $S_{\mathrm{bh}}$ that can give different results
\cite{DMB,CM2}, so for any of the formulas to be meaningful, the
quantities in them should be precisely defined.  Eqs. (\ref{eq:22}) and
(\ref{eq:23}) appear to refer to the logarithm of the number of black
hole states with a certain value of the horizon area $A =
4S_{\mathrm{BH}}$, or perhaps of states with the horizon area within a
certain narrow range, and so they might be called formulas for the
microcanonical entropy of a black hole.  Other formulas are for the von
Neumann entropy,
 \begin{equation}
 S_{\mathrm{vN}} = - \mathrm{Tr}(\rho \ln{\rho}),
 \label{eq:24}
 \end{equation}
for a density matrix $\rho$ representing a quantum ensemble of black
hole states, such as the canonical ensemble or the grand canonical
ensemble.

Both of these definitions have certain problems associated with them. 

Consider first the microcanonical ensemble, in which one generally
assumes that the quantity being fixed (e.g., the horizon area) has a
discrete spectrum, with each possible value having a certain
degeneracy.  The fine-grained microcanonical entropy would then be
defined for each possible value of the quantity being fixed and would
be an entropy that is the logarithm of the degeneracy.  Then there
seems to be two possibilities for the fine-grained microcanonical
entropy:

First, it could be that all of the possible area values are
nondegenerate, in which case the fine-grained microcanonical entropy
would be trivially zero for each allowed value of the area or other
fixed quantity.  If the fixed quantity were the energy rather than the
area, I would expect that nearly all of the energy eigenstates would be
nondegenerate, so that in this case the fine-grained microcanonical
entropy would be trivial.

(Classically, a black hole in vacuum or in the presence of a
cosmological constant would have its horizon area determined by a
simple function of its energy and other conserved quantities, such as
angular momentum.  However, when one includes quantum corrections, I
would not expect the horizon area, even if it is well defined, to be
determined precisely by the same simple function of the energy and
other commuting conserved quantities.  Even if the area and energy are
both given by Hermitian operators that commute, I would not expect
their precise eigenvalues to be so simply related as they are
classically.  Furthermore, I would not even expect both operators, if
they exist, to commute, since I see no fundamental reason why a black
hole configuration with definite values for the energy and other
commuting conserved quantities should also have a definite value for
its horizon area.  Conversely, I see no reason why black holes with
definite horizon areas, if such can exist, should have definite
energies.  Therefore, even if it is possible in the correct theory of
quantum gravity to define an area operator similar to that used in the
quantum geometry approach, and even if the eigenvalues of that operator
are restricted to macroscopically distinct values with large
degeneracies for the eigenstates with those eigenvalues, I would still
think it likely that the energy eigenstates would nearly all be
nondegenerate, with exponentially tiny gaps between the eigenvalues.)

Second, it could be that some or all of the possible area values are
degenerate, in which case the fine-grained microcanonical entropy would
be the logarithm of an integer (the degeneracy) for each allowed value
of the area or other fixed quantity.  However, in this case, I do not
see any strong reason why the degeneracy, and hence the microcanonical
entropy, would not vary greatly between nearby possible values of the
area.  For example, in quantum geometry, the allowed values of the
horizon area are often taken to be proportional to an integer-weighted
sum of $\sqrt{j(j+1)}$ over spins $j$ (such that $2j$ is a positive
integer), and nearby different values of these weighted sums would
typically have significantly different degeneracies.  An exception
might be the alternative assumption that there is an equally-spaced
spectrum of areas, but to me this seems to be merely an {\it ad hoc}
proposal without any significant justification.

Therefore, for the microcanonical ensemble, to get a nonzero entropy
that did not vary wildly between nearby values of the area or other
quantity, it would seem that one would need to add up the degeneracies
over a range of areas about some midpoint, or otherwise do some
smoothing of the degeneracies.  If the range of areas needed for the
smoothing is small enough that the logarithm of the smoothed degeneracy
does not vary greatly as the midpoint of the range is varied by an
amount comparable to the width of the range, then the resulting
smoothed microcanonical entropy would be fairly insensitive to the
smoothing procedure.  For very large $A = 4S_{\mathrm{BH}}$, it would
have an ambiguous additive constant with a linear dependence upon the
logarithm of the smoothing range, so the coefficient of
$\ln{S_{\mathrm{BH}}}$ in an expansion in powers and logarithms of
$1/S_{\mathrm{BH}}$ would depend upon how the smoothing range depends
upon $S_{\mathrm{BH}}$, but it would be well defined once a definite
prescription for the smoothing range is given.

However, it is not clear that this is indeed the case in the quantum
geometry approach, since it would seem that one might need to smooth
over a range of areas with a width that is comparable with unity (the
Planck area) to suppress fluctuations in the degeneracy.  But then the
degeneracy summed over the range of areas would depend nonlinearly on
the width of the range and on how the summed degeneracy depends on the
midpoint of the range, so it might be the case that the coefficient of
the logarithmic term in the smoothed microcanonical entropy would have
a more significant dependence upon the smoothing procedure.  Maybe it
would not, but that would remain to be checked.

Another problem I see with the microcanonical entropy is that at present
it is being calculated with the horizon area fixed, but it would seem to
me that that quantity would be difficult to measure.  Although quantum
geometry gives an operator for it, I am not convinced that it would be a
well-defined quantity in other approaches to quantum gravity (e.g., in
string/M theory).

It would seem to me much better to use something like the total energy
as the quantity to be fixed in a microcanonical ensemble.  This would
require appropriate asymptotic boundary conditions, such as
asymptotically anti-de Sitter boundary conditions with reflecting
boundary conditions at infinity \cite{HawPage}.  It would not work in
asymptotically flat spacetime, since then the infinite volume of phase
space at any finite energy would mean that it is entropically favorable
for the black hole to evaporate away, and so the microcanonical
ensemble would be dominated by diffuse radiation states rather than by
black hole states.

Even in asymptotically anti-de Sitter spacetime, small black holes would
decay away, so for black holes to dominate the microcanonical ensemble,
one needs the total energy $E$ to be large enough that the black hole
can be in stable equilibrium with the Hawking radiation surrounding
it.  In $d=4$ dimensions of spacetime, the condition for black holes to
dominate the microcanonical ensemble is given by Eq. (4.14) of
\cite{HawPage}, except that there is no upper limit for the energy if
one does not also ask, as we did there, that pure thermal radiation
without a black hole also be a local equilibrium state.

In a general dimension $d$ of spacetime, with $\ell \gg 1$ being the
length scale of the anti-de Sitter spacetime in Planck units (the length
we called $b$ in \cite{HawPage}), a black hole with horizon radius $r_+
\ll \ell$ would have a mass $M \sim r_+^{d-3}$, a temperature $T \sim
1/r_+$, and an entropy $S_{\mathrm{bh}} \sim r_+^{d-2} \sim
M^{(d-2)/(d-3)}$ in Planck units, omitting the dimension-dependent
numerical coefficients that would be of order unity for dimensions of
order unity.  The black hole would be surrounded by thermal Hawking
radiation of massless fields with energy density $\rho \sim T^d$ and
entropy density $s \sim T^{d-1}$ out to $r \sim \ell$, and then with
the rapidly rising anti-de Sitter gravitational potential causing the
local redshifted temperature, energy density, and entropy density all
to drop off exponentially with proper distance so that the total
radiation energy $E_{\mathrm{rad}} \sim \ell^{d-1} T^d$ and the total
radiation entropy $S_{\mathrm{rad}} \sim \ell^{d-1} T^{d-1} \sim (\ell
E_{\mathrm{rad}})^{1-1/d}$ are finite, even when one integrates over
the infinite proper spatial volume of the anti-de Sitter space.  (This
ignores the divergence one gets in integrating locally thermal
radiation down to the black hole horizon, where the local temperature
blueshifts to infinity.  Alternatively, one can assume that this
divergence at the horizon is regulated in some manner.)

Now the idea \cite{Haw3,PagGRG,HawPage} is that for fixed total energy
$E \approx M + E_{\mathrm{rad}}$, the equilibrium configuration has the
energy divided between the black hole mass $M$ and the radiation energy
$E_{\mathrm{rad}}$ in such a way to maximize the total entropy $S =
S_{\mathrm{bh}} + S_{\mathrm{rad}} \sim M^{(d-2)/(d-3)} +
[\ell(E-M)]^{(d-1)/d}$.  (I am ignoring the interaction energy between
the black hole and the radiation, which would be small if the black
hole is small and hence takes up only a tiny fraction of the volume
available to the radiation where the gravitational potential is not too
large.  However, it would not necessarily be negligible if one were
trying to calculate all terms in the total entropy $S$ that are large
in comparison with unity, so that one can get the number of states,
$\sim e^S$, correct with a fractional error of the order of unity.) 

If $E$ is too small, the only maximum of the entropy $S(M)$ would be at
$M=0$, meaning that pure thermal radiation with no black hole would
dominate the microcanonical ensemble.  For $E$ larger than some energy
$E_0$ that is some dimensionless number $c_0$ (depending on the number
$d$ of spacetime dimensions and on the radiation constant that depends
on the number and types of massless fields) times
$\ell^{(d-1)(d-3)/(2d-3)}$, two new extrema of $S(M)$ appear at
positive $M$, where the black hole temperature is the same at that of
the surrounding radiation.  When one perturbs either extremum by making
the black hole mass smaller and the radiation energy larger, both the
black hole and the surrounding radiation get hotter (since the black
hole with $r_+ \ll \ell$ has negative specific heat, and since the
radiation has positive specific heat).

However, for the extremum with smaller positive $M$, when one perturbs
to yet smaller $M$, the black hole temperature rises more than the
surrounding radiation, so the black hole stays hotter than the
surrounding radiation and is unstable to evaporating away. 
Alternatively, if the black hole mass $M$ is perturbed to a larger
value than the extremum, the black hole temperature cools more than the
surrounding radiation, so then the black hole is unstable to absorbing
yet more radiation and continuing to grow.  This extremum is thus a
local minimum of $S(M)$ and represents unstable equilibrium between a
black hole and surrounding radiation with the same temperature.

On the other hand, for the extremum with larger positive $M$, when one
perturbs $M$ to a smaller value, the black hole temperature rises less
than the temperature of the radiation, so the radiation gives a
negative feedback on the shrinkage of the hole, keeping it stable. 
Alternatively, if the black hole mass $M$ is perturbed to a larger
value than the extremum, the black hole temperature cools less than the
surrounding radiation, so then the black hole is hotter than the
surrounding radiation and thus emits its excess energy to return to its
equilibrium mass.  This extremum is thus a local maximum of $S(M)$ and
represents a locally stable equilibrium between a black hole and
surrounding radiation with the same temperature.

Note that if one started with a black hole at the unstable extremum
(the smaller positive value of $M$ that gives an extremum) and then
perturbed $M$ to a larger value, at first the black hole would have an
unstable growth as described above, but eventually it would settle down
at the locally stable extremum with the larger value of $M$.  On the
other hand, if one perturbed the unstable equilibrium to give $M$ a
smaller value, the black hole would continue to shrink until it
disappeared, giving the other locally stable configuration, pure
thermal radiation.  It would require a nonperturbative fluctuation to
transit between these two locally stable configurations.

Since the Hawking temperature for a black hole much smaller than the
characteristic length scale $\ell$ of the anti-de Sitter spacetime of
$d$ spacetime dimensions (black hole horizon radius $r_+ \ll \ell$) is
proportional to $M^{-1/(d-3)}$, and since the temperature of thermal
radiation in $d$ dimensions is proportional to
$E_{\mathrm{rad}}^{1/d}$, one can readily see that the condition for
the black hole temperature to rise less than the radiation when a small
amount of energy is transferred from the black hole (reducing $M$) to
the radiation (increasing $E_{\mathrm{rad}} \approx E-M$ by the
essentially the same amount), starting at the condition when the two
temperatures are equal, is that $E_{\mathrm{rad}} < (1-3/d)M$, or
$E_{\mathrm{rad}} < [(d-3)/(2d-3)]E$, or $M > [d/(2d-3)]E$.  When one
saturates these inequalities and requires that the black hole and
radiation temperatures be equal, one gets $E = E_0 = c_0
\ell^{(d-1)(d-3)/(2d-3)}$ and can evaluate the numerical coefficient
$c_0$ for various spacetime dimensions and values of the radiation
constant \cite{Haw3,PagGRG,HawPage}.

Therefore, for $E > E_0$, there is a locally stable equilibrium
microcanonical configuration with a black hole having positive mass
$M$. However, the pure thermal radiation configuration with $M=0$ also
remains a locally stable equilibrium microcanonical configuration, at
least for $E < E_2 = c_2 \ell^{d-3} \sim \ell^{d-3} \gg E_0$ with some
numerical coefficient depending purely upon $d$ (assuming that the
radiation is purely massless radiation).  (For $E > E_2$, any
configuration of pure radiation would be dynamically unstable to
collapsing to form a black hole, e.g., by the Jeans instability.) 
Which of the two locally stable configurations will dominate the
microcanonical ensemble depends upon their entropies.

For $E_0 < E < E_1 = c_1 \ell^{(d-1)(d-3)/(2d-3)}$ with another
numerical coefficient $c_1$ that is somewhat larger than $c_0$ (e.g.,
larger by a factor of about 1.314 for $d=4$ \cite{HawPage}), the pure
thermal radiation configuration has more entropy and so dominates.  But
for $E > E_1$, the configuration with a black hole in locally stable
thermal equilibrium with surrounding radiation having less energy than
$[(d-3)/d]M$ has more entropy and so dominates the microcanonical
ensemble.

Therefore, we can say that for $E > E_1 \sim \ell^{(d-1)(d-3)/(2d-3)}$,
black hole configurations dominate the microcanonical ensemble of fixed
energy $E$ in asymptotically anti-de Sitter spacetime.

One would expect that in a quantum analysis, one would get a discrete
set of energy eigenvalues.  In principle, from sufficient (future)
knowledge of the quantum theory of gravity and matter in asymptotically
anti-de Sitter spacetime with length scale $\ell$, one could count the
number $N(E)$ of energy eigenstates up to any energy $E$, and then one
could use this to define various microcanonical entropies.  Again one
would face the challenge that the fine-scale microcanonical entropy,
the logarithm of the degeneracy of a precise energy eigenvalue, is
likely either to be zero (if the energy eigenvalues are nondegenerate,
as I would expect, unless perhaps $\ell$ can be tuned to give a
degeneracy) or else fluctuate between nearby energy eigenvalues (if
various ones are degenerate).

Therefore, it seems likely that some smoothing over nearby energy
eigenvalues would be needed.  However, since for $E \gg E_1$ a
semiclassical analysis, using the Bekenstein-Hawking entropy for the
dominant black hole with $M \approx E$, suggests that, very crudely,
$N(E) \sim \exp{[c E^{(d-2)/(d-3)}]}$ for some readily-calculable
numerical coefficient $c$ that depends purely upon $d$.  Thus $N(E)$
rises very rapidly with $E$ (faster than exponentially with the
energy).  Unless the energy eigenvalues are clumped into values with
huge degeneracies that are then separated by macroscopic amounts (e.g.,
by amounts comparable with the black hole temperature), which seems
very implausible to me, one would expect a huge number of eigenvalues
within any range of energies that is not extraordinarily narrow. 
Therefore, one could do a fairly generic smoothing over any range of
energies (but a range large enough to contain a huge number of energy
eigenvalues) to get a coarse-grained microcanonical entropy.

However, there would remain some ambiguities in the coarse-grained
microcanonical entropy, depending upon how the range of energies was
defined.  One of the least {\it ad hoc} definitions would be to define
 \begin{equation}
 S_{\mathrm{microcanonical}}(E) = \ln{N(E)},
 \label{eq:25}
 \end{equation}
where $N(E)$ counts all of the energy eigenstates up to energy $E$, so
then the range would go all the way from the minimum energy state to
$E$.

Nevertheless, one could alternatively define coarse-grained
microcanonical entropies with narrower ranges, say
 \begin{equation}
 S'_{\mathrm{microcanonical}}(E) = \ln{[N(E+\Delta)-N(E-\Delta)]},
 \label{eq:26}
 \end{equation}
the logarithm of the number of energy eigenstates with energy within
$\Delta$ of $E$.  If $N(E+\Delta)-N(E-\Delta) \ll N(E-\Delta)$, which
for the rapidly growing $N(E)$ implies $\Delta \ll E$ but is much
stronger than that condition, $S'_{\mathrm{microcanonical}}(E)$ would
have a roughly linear dependence upon the logarithm of $\Delta/E$.

For example, if for the sake of argument one assumed that $N(E) =
\exp{(cE^\alpha)}$ for some constants $c$ and $\alpha$ (e.g., for
$\alpha = (d-2)/(d-3)$), here and henceforth neglecting the truncation
to the largest integer not greater than $\exp{(cE^\alpha)}$ that would
be needed for $N(E)$ to be an integer, then
 \begin{equation}
 S \equiv S_{\mathrm{microcanonical}}(E) = \ln{N(E)} = cE^\alpha,
 \label{eq:27}
 \end{equation}
whereas for $c\alpha E^{\alpha-1} \Delta \ll 1$ so that
$N(E+\Delta)-N(E-\Delta) \ll N(E-\Delta)$, one gets
 \begin{eqnarray}
 S' &\equiv& S'_{\mathrm{microcanonical}}(E)
    = \ln{[N(E+\Delta)-N(E-\Delta)]} \nonumber \\
    &\approx& \ln{[e^{cE^\alpha}(e^{c\alpha E^{\alpha-1}\Delta}
                               - e^{-c\alpha E^{\alpha-1}\Delta})]}
    \approx \ln{[e^{cE^\alpha} 2c\alpha E^{\alpha-1} \Delta]}
    \nonumber \\
    &=& S + \ln{S} + \ln{(\Delta/E)} + \ln{(2\alpha)}.
 \label{eq:28}
 \end{eqnarray}
 
This example shows how logarithms can come in when one goes from one
coarse graining to another.  Here the coefficient of the logarithm of
the energy depends on how the width of the range depends on the energy. 
For example, if we set $\Delta = \beta E^\gamma$, then
 \begin{equation}
 S' = S + (\alpha+\gamma-1)\ln{E} + \ln{(2\alpha\beta c)}
    = S + {\alpha+\gamma-1 \over \alpha} \ln{S} + \ln{(2\alpha\beta)}
        - {\gamma-1\over\alpha} \ln{c}.
 \label{eq:29}
 \end{equation}
Thus by choosing $\gamma$ appropriately, one can get any coefficient of
the logarithm term that one wants.  In this case it would be absent if
$\gamma = 1-\alpha$, which would be the case if the range were made to
give equal widths for the classical horizon area $A \propto E^\alpha$,
rather than equal widths for any other power of $E$.

Thus we see that the coefficient of a logarithm term in an entropy does
not have any meaning unless one carefully specifies how the entropy is
to be defined.

What would be more meaningful would be a difference between an entropy
calculated classically (or semiclassically) and one calculated with the
full theory of quantum gravity and some specific definition of the
entropy.  Ideally, one would like both entropies to be defined the same
way, but since a classical or semiclassical calculation is not likely
to lead to discrete energy eigenstates, there may not be a precise
classical or semiclassical analogue for many of the definitions of
entropy possible in the quantum theory with definite discrete energy
eigenstates.

The classical or semiclassical calculation would be easiest if one
fixed the black hole area, or a range of areas, since the classical
Bekenstein-Hawking formula for the entropy is simply one-fourth the
area.  However, as mentioned above, for a comparison with a quantum
theory, it would probably be better to do a comparison in terms of a
quantity like the energy that is more likely to be well defined in the
quantum theory (under suitable conditions).  For example, in string
theory on an asymptotically anti-de Sitter background, presumably the
energy would be well defined, but it may be doubtful whether any
horizon or any horizon area is well defined.

If one used energy rather than horizon area as the quantity that may be
well defined both classically and quantum mechanically, then one would
like to compare some quantum entropy with a classical calculation for
it.  For example, one might take the classical entropy to be the
Bekenstein-Hawking entropy $S_{\mathrm{BH}} \equiv A/4$ for a classical
configuration with the same energy $E$.

However, if the semiclassical description is of a black hole surrounded
by thermal radiation, one would expect the black hole entropy to be
augmented by the entropy of the radiation (at least the part of the
radiation far from the hole, if the black hole entropy includes the
effects of radiation very near the horizon).  One might want to include
some approximation for this entropy within the semiclassical entropy. 
But then it becomes a question of precisely how this is to be done
within the semiclassical approximation, if one wants to compare it with
the quantum result (once that is known, with some suitable definition of
the quantum entropy from the spectrum of energy eigenstates).

For example, above I alluded to an approximation in which one took a
small black hole in the vacuum Schwarzschild-anti-de Sitter spacetime
(which classically has well-defined relations between the black hole 
horizon radius $r_+$, the mass $M$, the temperature $T$, and the
Bekenstein-Hawking entropy $S_{\mathrm{BH}} \equiv A/4$ for a given
anti-de Sitter length scale $\ell$) and then just added on the energy
and entropy of thermal radiation in anti-de Sitter spacetime (for this
ignoring the change in the geometry from the black hole).

Even for the latter energy and entropy, there is some ambiguity on how
it is to be calculated classically or semiclassically.  The simplest
approximation would be to assume that the massless radiation had local
energy and entropy densities given by the flat spacetime radiation
constant multiplied by the appropriate dimension-dependent powers of
the local temperature, and to neglect the back reaction of the
radiation on the geometry.  A slightly improved calculation would be to
take the quantum canonical ensemble of the massless radiation fields in
the fixed anti-de Sitter background and calculate from it the
expectation value of the energy and the von Neumann entropy.  A further
improved but difficult calculation would be to take the quantum
canonical ensemble for the radiation fields in a static classical
asymptotically anti-de Sitter background (with no black hole at this
stage) that solves the Einstein equation with the cosmological constant
and the expectation value of the stress-energy tensor of the quantum
fields in their canonical ensemble.

However, when one has both the black hole and the radiation, one would
think that the semiclassical calculation should be improved by
including the effect of the black hole on the asymptotically
anti-de Sitter metric where the thermal radiation resides.  A classical
way to attempt to do this would be to take the
Schwarzschild-anti-de Sitter metric for the black hole with the negative
cosmological constant and assume that the radiation is described by a
traceless stress-energy tensor that has isotropic pressures and an
energy density given by the flat spacetime radiation constant
multiplied by the appropriate power of the local Hawking temperature. 
But this would lead to a divergence of the energy and entropy densities
of the radiation at the black hole horizon, where the local Hawking
temperature diverges.  Therefore, it would be better (though more
difficult) to take the canonical ensemble for the quantum radiation
field in the Schwarzschild-anti-de Sitter metric, with the divergence at
the horizon regulated.  An even better but much more difficult
calculation would be to take the quantum canonical ensemble for the
radiation fields in a static classical asymptotically anti-de Sitter
background with the black hole present that is solves the Einstein
equation with the cosmological constant and the expectation value of
the stress-energy tensor of the quantum fields in their canonical
ensemble.  If this gives an unambiguous entropy for the quantum fields,
one could add on one-quarter of the black hole area to get a
semiclassical entropy as a function of the ADM energy of the
configuration.

Then one could compare this semiclassical result with the full quantum
gravity result once that is known.  However, the later will depend on
how the entropy is defined, so it still would not be unambiguous what
difference quantum gravity makes to the entropy.  But at least one
would presumably get a definite result for each possible definition of
the quantum entropy.

If one is just looking for a difference involving the logarithm of the
Bekenstein-Hawking entropy $S_{\mathrm{BH}} \equiv A/4$, one would not
need the precise semiclassical or quantum entropy, but only the terms
that involve positive powers or logarithms of $S_{\mathrm{BH}}$. 
However, even this might be difficult for the semiclassical entropy.

For example, if we consider the microcanonical ensemble discussed above
when $E$ is comparable with $E_1$, one could presumably write the
semiclassical entropy as a power series (perhaps including logarithms)
in the small classical quantity $r_+/\ell$, the linear size of the
horizon divided by the linear characteristic size of the asymptotically
anti-de Sitter spacetime.  However, for $E \sim E_1 \sim
\ell^{(d-1)(d-3)/(2d-3)}$, one has $r_+/\ell \sim
\ell^{-(d-2)/(2d-3)}$, and the leading classical term in the entropy
goes as $E^{(d-2)/(d-3)} \sim r_+^{d-2} \sim \ell^{(d-1)(d-2)/(2d-3)}
\sim (r_+/\ell)^{-(d-1)}$.  Therefore, if this leading term is
multiplied by a correction factor that is an increasing power series in
$r_+/\ell$ (starting with unity), and if all positive powers occurred,
one would need the leading classical term and $d-1$ others to include
all terms in the entropy that are not small for small $r_+/\ell$.  (The
term that is zero-order in $r_+/\ell$, in the product of the leading
term and the correction factor power series, might be a logarithmic
term.)  Thus we might need to calculate $d$ terms in a power series to
determine the semiclassical entropy (as a function of the ADM mass) to
enough precision to identify a difference with the quantum entropy
(once that more difficult calculation can be performed) that is
logarithmic in $S_{\mathrm{BH}}$.

For example, there could be terms in the power series in $r_+/\ell$
that would express the effect of the black hole on the radiation, the
self-gravitational effects of the radiation itself, departures of the
expectation value of the stress-energy tensor of the radiation from
having an isotropic pressure, interactions between these effects, etc.

One way to circumvent some of these problems with the semiclassical
treatment of the radiation would be to take a black hole in
asymptotically anti-de Sitter spacetime that is very large in
comparison with the anti-de Sitter length scale, so $r_+ \gg \ell$. 
Then the exponentially rising gravitational potential at distances much
greater than $\ell$ from the black hole would suppress the thermal
radiation so that it would have very little energy and entropy in
comparison with that of the hole (at least if one regulated the
divergence at the black hole horizon).  Then in the limit that
$r_+/\ell$ is made very large (and also $\ell$ is kept large in Planck
units to keep the geometry nearly classical), one might be able to
ignore the effects of radiation altogether on the semiclassical
calculation of the energy and entropy and simply take their values from
the Schwarzschild-anti-de Sitter metric and the Bekenstein-Hawking
entropy.  It would be the relationship between these two quantities
that one could use to compare with the quantum entropy.

Another advantage of using $r_+ \gg \ell$ is that then not only the
microcanonical ensemble but also the canonical ensemble is well defined
in asymptotically anti-de Sitter boundary conditions \cite{HawPage}. 
(In asymptotically flat spacetime, the canonical ensemble is not
defined, because the number of states of a black hole rises
indefinitely faster than exponentially with the energy \cite{Haw3}.) 
Therefore, not only is there the natural coarse-grained microcanonical
entropy defined by Eq. (\ref{eq:27}), the logarithm of the number of
energy eigenstates with energy below $E$,
$S_{\mathrm{microcanonical}}(E) = \ln{N(E)}$, but there is also the
canonical entropy that may be defined to be a function of the
energy expectation value $\langle E \rangle$:
 \begin{equation}
 S_{\mathrm{canonical}}(\langle E \rangle)
  = \beta \langle E \rangle + \ln{Z},
 \label{eq:30}
 \end{equation}
where the partition function is
 \begin{equation}
 Z(\beta) = \sum_i e^{-\beta E_i},
 \label{eq:31}
 \end{equation}
with the index $i$ running over all energy eigenstates with energy $E_i$
(more than one of which could be the same if any energy eigenvalues are
degenerate), and where
 \begin{equation}
 \langle E \rangle = - {\partial\over \partial\beta} \ln{Z}
 \label{eq:32}
 \end{equation}
denotes the expectation value of the energy.

For energies such that $N(E)$ is very large, one can approximate the
discrete function $N(E)$ (taking only integer values) with a continuous
function, say $n(E) = e^{s(E)}$ with $s(E)$ being some continuum
approximation to $S_{\mathrm{microcanonical}}(E) = \ln{N(E)}$.  Then
the sum above for the partition function $Z(\beta)$ would be replaced
by the integral
 \begin{equation}
 Z(\beta) \approx z(\beta) = \int e^{-\beta E} dn
  = \int e^{s-\beta E(s)} ds,
 \label{eq:33}
 \end{equation}
where $E(s)$ is the inverse function to $s(E) = \ln{n(E)}$.

Now define the auxiliary function $T(s) \equiv dE/ds$ (a microcanonical
temperature) and assume that it is always positive and, at least for
all $s$ above some value $s_0$ where it has a local minimum value
$T_0$, that $T(s)$ rises smoothly, monotonically, and indefinitely with
$s$.  This implies that for $s > s_0$, the exponent in the integral,
$I(s) = s-\beta E(s)$, is concave downward, with $dI/ds = 1 - \beta
T(s)$ decreasing monotonically with $s$.  Therefore, for any fixed
$\beta < 1/T_0$, there will be a unique extremum of $I(s)$, at the
unique value of $s = s(\beta) > s_0$ where $T(s(\beta)) = 1/\beta$. 
Let this extremum value of $I(s)$ be denoted $I_{\mathrm{max}}(\beta) =
s(\beta) - \beta E(\beta)$, with $E(\beta) = E(s(\beta))$ being the
value of $E$ at $s = s(\beta)$.

It is also convenient to define the microcanonical heat capacity
function $C(s) = dE/dT$, which will be positive for $s > s_0$, since
then both $E$ and $T$ are rising monotonically with $s$.  Let its value
at the extremum value of $I(s)$ be $C(\beta) = C(s(\beta))$.  Since one
can readily show that $d^2I(s)/ds^2 = -1/C(s)$, for $s$ sufficiently
near the value $s(\beta)$ that extremizes $I(s)$, one has
 \begin{equation}
 I(s) \approx I_{\mathrm{max}}(\beta) - {1\over 2C(\beta)}[s-s(\beta)]^2.
 \label{eq:34}
 \end{equation}

If this approximation is valid over the dominant part of the integral
(\ref{eq:33}) for $z(\beta)$, one can do the integral by the
saddle-point approximation and get the value
 \begin{equation}
 Z(\beta) \approx z(\beta) \approx \sqrt{2\pi C(\beta)}\ 
 e^{I_{\mathrm{max}}(\beta)}.
 \label{eq:35}
 \end{equation}

Then one can use Eqs. (\ref{eq:32}) and (\ref{eq:30}) to calculate
 \begin{equation}
 \langle E \rangle \approx E(\beta) + {T^2\over 2C}{dC\over dT},
 \label{eq:36}
 \end{equation}
where $T$ and $C$ and $dC/dT$ are evaluated at $s = s(\beta)$, and to
calculate
 \begin{equation}
 S_{\mathrm{canonical}} \approx s + {1\over 2}\ln{(2\pi C)}
     + {T\over 2C}{dC\over dT},
 \label{eq:37}
 \end{equation}
where again everything is evaluated at $s = s(\beta)$.

For a large black hole in anti-de Sitter spacetime, asymptotically at
large $r_+/\ell$ one has $E \approx M \sim r_+^{d-1}/\ell^2$ and $s
\approx S \sim r_+^{d-2} \sim E^\alpha$ with $\alpha = (d-2)/(d-1)$, so
$E(s) \sim s^{(d-1)/(d-2)}$, $T(s) = dE/ds \sim s^{1/(d-2)}$, and $C(s)
= dE/dT \approx (d-2)s$.  Therefore, one gets
 \begin{equation}
 S_{\mathrm{canonical}} \approx s + {1\over 2}\ln{s} + O(1).
 \label{eq:38}
 \end{equation}

Remembering that $s$ is a continuum approximation to
$S_{\mathrm{microcanonical}}$, one gets that when either entropy is
large, these two entropies differ mainly by the logarithm of the
entropy (dropping terms of the order of unity):
 \begin{equation}
  S_{\mathrm{canonical}} \approx S_{\mathrm{microcanonical}}
   + {1\over 2}\ln{S_{\mathrm{microcanonical}}},
 \label{eq:39} \end{equation}
 \begin{equation}
  S_{\mathrm{microcanonical}} \approx S_{\mathrm{canonical}}
   - {1\over 2}\ln{S_{\mathrm{canonical}}}.
 \label{eq:40}
  \end{equation}

Similar relations have been given in \cite{DMB,CM2}, though they used a
microcanonical entropy defined as the logarithm of the density of
states as a function of energy, rather than my use of the
microcanonical entropy as the logarithm of the total number of states
of lower energy.  From Eqs. (\ref{eq:28}) and (\ref{eq:29}) above, one
can see that one can get further changes in the coefficient of the
logarithm by going to a microcanonical entropy defined as the logarithm
of the number of states within some range of energies, depending on how
the range varies with the midpoint of the range.

Therefore, entropies need to be defined carefully before there is any
unambiguous meaning to logarithmic corrections.  But even if one does
choose a precise definition, it may be a long time before we have
sufficient knowledge of quantum gravity to be able to calculate the
correct answers for the logarithmic terms in the entropy.

\section{Conclusions}

Black holes are perhaps the most highly thermal objects in the
universe (though they are very cold for stellar mass black holes).
Their phenomenological thermodynamic properties are very well
understood (at least for quasistationary semiclassical black holes),
but a good understanding of their microscopic degrees of freedom is
lacking.  Although it seems that black holes are rather like other
thermal objects (say in having such degrees of freedom that carry the
information imparted into them and restore this information to the
outer universe when the black holes evaporate away), one is not yet
completely sure that this is the case, or, if it is, where and how the
microscopic degrees of freedom store the information.  Therefore,
although we have gained an enormous amount of information about black
holes and their thermal properties in the past thirty years, it seems
that there is even much more that we have yet to learn.

Spacetime limitations on the author have prevented this review from
being anywhere near complete.  For other recent (and often more nearly
complete) reviews, see
\cite{Frolov96,Wald97,Sor3,Jacobson97,FF,Mukohyama98,Kiefer99,
Jacobson99,Mukohyama99,Wald99,Majumdar0,Majumdar1,Kiefer02,
Israel03,Jacobson03,Damour,Das,Fursaev}.

\section*{Acknowledgments}

I am grateful for many discussions with, among others, Abhay Ashtekar,
Tom Banks, Andrei Barvinsky, Jacob Bekenstein, David Boulware, Raphael
Bousso, Steve Carlip, Brandon Carter, Paul Davies, David Deutsch, Bryce
DeWitt, Doug Eardley, Richard Feynman, Willy Fischler, Eanna Flanagan,
Larry Ford, Valeri Frolov, Ulrich Gerlach, Gary Gibbons, Jim Hartle,
Stephen Hawking, Gerard 't Hooft, Gary Horowitz, Werner Israel, Juan
Maldacena, Robb Mann, Don Marolf, Charlie Misner, Rob Myers, Ted
Newman, Igor Novikov, Leonard Parker, Amanda Peet, Roger Penrose, Joe
Polchinski, Chris Pope, Bill Press, Richard Price, Rafael Sorkin, Mark
Srednicki, Alexie Starobinsky, Andy Strominger, Lenny Susskind, Saul
Teukolsky, Kip Thorne, Bill Unruh, Cumrun Vafa, Bob Wald, John Wheeler,
Ed Witten, Bill Wootters, Yakov Zel'dovich, Andrei Zelnikov, and
Wojciech Zurek.  I am grateful for referees for pointing out the
relevance of references \cite{Par1,Par2,Par3,Par4,Full1,Full2} and for
encouraging me to include a discussion of logarithmic corrections to
the entropy of a black hole.  Financial support has been provided by
the Natural Sciences and Engineering Research Council of Canada.


\end{document}